\documentclass[aps,prb,twocolumn,showpacs,byrevtex]{revtex4}

\usepackage{times,mathptm}
\usepackage[dvips]{graphicx,color}
\newcommand{\bqa}{\begin{eqnarray*}}
\newcommand{\eqa}{\end{eqnarray*}}

\begin{document}

\title{Fluorine-induced local magnetic moment in graphene: A hybrid DFT study}
\author{Hyun-Jung Kim and Jun-Hyung Cho$^*$}
\affiliation{Department of Physics and Research Institute for Natural Sciences, Hanyang University,
17 Haengdang-Dong, Seongdong-Ku, Seoul 133-791, Korea
}
\date{\today}

\begin{abstract}
Recent experimental evidence that fluorinated graphene creates local magnetic moments around F adatoms has not been supported by semilocal density-functional theory (DFT) calculations where the adsorption of an isolated F adatom induces no magnetic moment in graphene. Here, we show that such an incorrect prediction of the nonmagnetic ground state is due to the self-interaction error inherent in semilocal exchange-correlation functionals. The present hybrid DFT calculation for an isolated F adatom on graphene predicts not only a spin-polarized ground state with a spin moment of ${\sim}$1 ${\mu}_{\rm B}$, but also a long-range spin polarization caused by the bipartite nature of the graphene lattice as well as the induced spin polarization of the graphene states. The results provide support for the experimental observations of local magnetic moments in fluorinated graphene.
\end{abstract}


\maketitle


Control of the electronic and magnetic properties of graphene is of significant technological interest in emerging spintronics applications such as new types of magnetic sensors as well as spin-based random access memory.~\cite{tombros,castro1} In particuler, the adsorption of adatoms on graphene has been utilized to change the unique electronic band structure of graphene, offering an effective way to create local magnetic moments in nonmagnetic graphene.~\cite{yazyev1,castro2,uch} Such a chemical functionalization of graphene with adatoms can thus be useful to manipulate spin transport in graphene, which is essential to the operation of spintronics devices. For this respect, there have been a number of theoretical and experimental studies to explore adatom-induced magnetism in graphene.~\cite{creary,sofo1,yazyev2,bouk,sahin,osuna,nair,hong1,hong2,santos,sofo2,liu} However, the precise identification and characterization of such magnetism in graphene are still at an early stage and sometimes remain controversial. For instance, experimental and theoretical studies of fluorinated graphene have reported inconsistent results.~\cite{nair,hong1,hong2,sofo2,santos,liu}

Recently, Hong $et$ $al$.~\cite{hong1} observed a colossal negative magnetoresistance in dilute fluorinated graphene, suggesting adatom-induced magnetism. Subsequently, they observed an anomalous saturation of the phase coherence length in the same system, which provided evidence for spin-flip scattering caused by an adatom-induced local magnetic moment.~\cite{hong2} By magnetization measurements as a function of externally applied magnetic field, Nair $et$ $al$.~\cite{nair} observed spin-1/2 paramagnetism in the fluorinated graphene (CF$_x$) samples with $x$ increasing from 0.1 to 1.0. Since the measured number of paramagnetic centers is three orders of magnitude less than the measured number of F adatoms on graphene,~\cite{nair} they interpreted the paramagnetic centers in terms of local magnetic moments arising from the edges of F clusters which are easily formed on graphene.~\cite{osuna,sahin} Therefore, these recent experiments~\cite{hong1,hong2,nair} evidenced the presence of adatom-induced local magnetic moments in fluorinated graphene with a wide range of F concentration.

In contrast to the above-mentioned experimental~\cite{hong1,hong2,nair} evidences of F-induced local magnetic moment in graphene, previous DFT studies with semilocal treatment of exchange and correlation, i.e., generalized gradient approximation (GGA), predicted that an isolated F adatom on graphene has a nonmagnetic ground state.~\cite{liu,santos,sofo2} According to scanning tunneling microscopy experiments, a single-atom defect on graphene gives rise to a quasilocalized state at the Fermi level, which extends over several nanometers around the defect.~\cite{kelly,hong1} We note that semilocal GGA functionals tend to stabilize artificially delocalized electronic states due to their inherent self-interaction error (SIE),~\cite{sie1,sie2,sie3} since delocalization reduces the spurious self-repulsion of electron itself. In this sense, previous~\cite{liu,santos,sofo2} GGA calculations for an isolated F adatom on graphene seem to overestimate the relative stability of a rather delocalized non-magnetic configuration compared to a spin-polarized configuration, resulting in a nonmagnetic ground state. Therefore, it is interesting to examine if the observed~\cite{hong1,hong2,nair} local magnetic moment in fluorinated graphene can be predicted by the correction of the SIE with an exchange-correlation functional beyond the GGA.

In this Rapid Communication, we present a new theoretical study for an isolated F adatom on graphene, which extends the previous work by considering a hybrid exchange-correlation functional. Unlike previous GGA calculations predicting a nonmagnetic ground state,~\cite{santos,sofo2,liu} the present hybrid DFT calculation predicts that an isolated F adatom induces a spin moment of ${\sim}$1 ${\mu}_{\rm B}$ in graphene. Especially, we find that F adsorption gives rise to a long-range spin polarization of $p_z$ orbitals in graphene as well as a band-gap opening of graphene ${\pi}$ bands. Our prediction of a spin-polarized ground state is consistent with recent experimental~\cite{hong1,hong2,nair} observations of F-induced local magnetic moments in graphene.

The present first-principles DFT calculations were performed using the FHI-aims code~\cite{aims-a} for an accurate, all-electron description based on numeric atom-centered orbitals, with ``tight" computational settings and accurate tier-2 basis sets. For the exchange-correlation energy, we employed both the GGA functional of Perdew-Burke-Ernzerhof (PBE)~\cite{pbe} and the hybrid functional of Heyd-Scuseria-Ernzerhof (HSE).~\cite{hse} For simulation of the adatom-graphene system, we used a periodic sheet geometry in which one adatom adsorbs on one side of graphene within a 3$\sqrt{3}$${\times}$3$\sqrt{3}R$30$^{\circ}$ unit cell (see Fig. 1) and adjacent graphene sheets are separated with a vacuum spacing of 50 ${\rm \AA}$. In order to remove the dipole interactions between periodic sheets, we applied corrections to the electrostatic potential and the total energy.~\cite{neu} For the Brillouin zone integration, we used 12${\times}$12 ${\bf k}$ points in the surface Brillouin zone. All the atoms were allowed to relax along the calculated forces until all the residual force components are less than 0.02 eV/{\AA}.

We begin to optimize the atomic structure of an isolated F adatom on graphene using the PBE functional. Here, we use our calculated graphene lattice constant of 2.467 {\AA}, which is very close to the experimental~\cite{elias} value of 2.46 {\AA}. The optimized structure of such an F-graphene system is displayed in Fig. 1. It is seen that the C atom (labeled as C$_0$) bonded to the F adatom shows a protrusion from the graphene layer, due to the change of bonding character from $sp_{2}$ to $sp_{3}$-like hybridization. The vertical displacement of C$_0$ from the graphene layer and the bond length $d_{\rm F-C_0}$ ($d_{\rm C_1-C_2}$) between the F (C$_0$) and C$_0$ (C$_1$) atoms are calculated to be 0.510 and 1.553 (1.482) {\AA}, in good agreement with previous~\cite{santos,sofo2,liu} PBE calculations. Figure 2(a) shows the calculated band structure for the F-graphene system. We find that the Fermi level $E_{\rm F}$ is shifted lower in energy relative to the Dirac point of graphene by 0.45 eV and the F-induced state lying at the Fermi level is partially occupied, indicating some charge transfer from the graphene ${\pi}$ states to the F-induced state. Indeed, our analysis of the Mulliken charges shows a charge transfer of 0.30 $e$ from graphene to the F adatom, representing a partial ionic bonding between the F adatom and graphene. We note that the charge character [see the inset of Fig. 2(a)] of the F-induced state represents a quasilocalized electron which is distributed dominantly over the sites of the sublattice (labeled ${\beta}$) opposite to the one (labeled ${\alpha}$) to which an isolated F adatom is attached.

\begin{figure}[tb]
\includegraphics[width=0.45\textwidth]{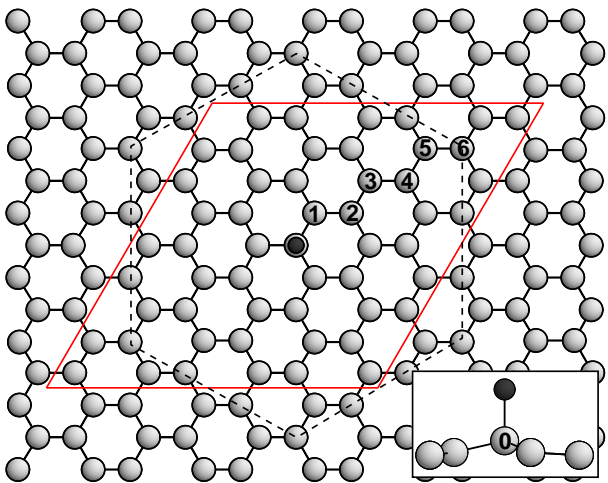}
\caption{(Color online) Top view of the optimized structure of the F-graphene system. Large (bright) and small (dark) circles represent C and F atoms, respectively. The inset highlights the F-adsorbed region with the side view. The unit cell and the Voronoi cell are shown as the solid and dashed lines, respectively. The numbers in C atoms represent the atomic site for the magnetic moment $m_i$ plotted in Fig. 3(c). }
\end{figure}

\begin{figure}[tb]
\includegraphics[width=0.45\textwidth]{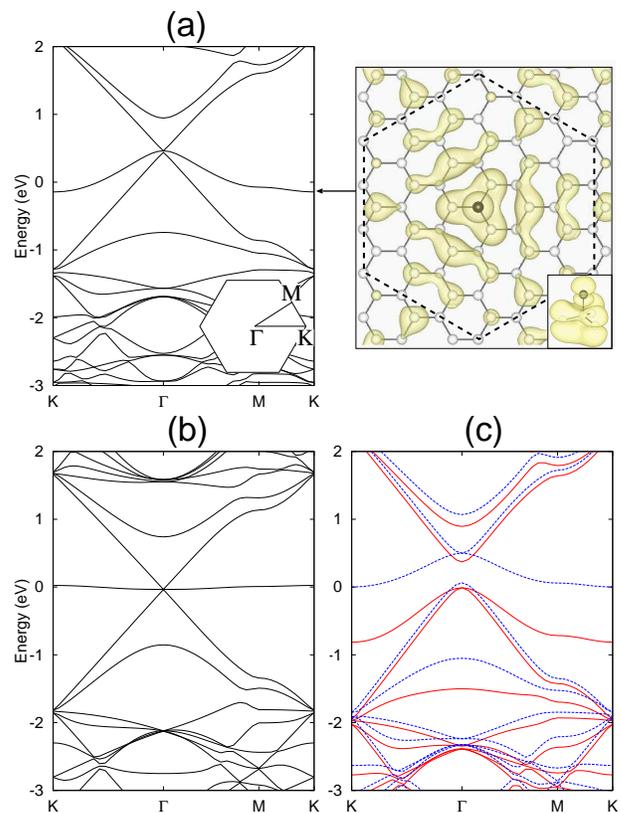}
\caption{(Color online) Spin-unpolarized band structures of the (a) F-graphene and (b) H-graphene systems computed using the PBE functional. The spin-polarized band structure of the F-graphene system computed using the HSE functional is given in (c). The band dispersions are plotted along the symmetry lines of the Brillouin zone of the 3$\sqrt{3}$${\times}$3$\sqrt{3}R$30$^{\circ}$ unit cell [see the inset in (a)]. The inset in (a) also shows the charge character of the F-induced state at the K point with an isosurface of 0.002 $e$/{\AA}$^3$, together with a perspective view around the F adatom. The energy zero represents the Fermi level. The majority and minority bands in (c) are drawn with the solid and dashed lines, respectively.}
\end{figure}

To support the recent experimental~\cite{hong1,hong2,nair} evidences that fluorinated graphene exhibits local magnetic moments, we perform the additional spin-polarized calculation for an isolated F adatom on graphene using the PBE functional. However, we are unable to find the stabilization of a spin-polarized configuration, consistent with previous GGA calculations.~\cite{liu,santos,sofo2} On the other hand, it is noticeable that our PBE calculation for an isolated H adatom on graphene (employing the same 3$\sqrt{3}$${\times}$3$\sqrt{3}R$30$^{\circ}$ unit cell as in the calculation of the F-graphene system) predicts a spin-polarized ground state with a magnetic moment of 1 ${\mu}_{\rm B}$ per adatom, in good agreement with previous GGA calculations.~\cite{santos,yazyev2,bouk,sofo1} Figure 2(b) shows the band structure of the spin-unpolarized solution for the H-graphene system. It is seen that the H-induced state has a very narrow band width of 0.06 eV, leading to a splitting of the spin-up and spin-down bands (see Fig. 1 of the Supplemental Material~\cite{supple}) via the Stoner-type ferromagnetic exchange~\cite{yazyev2,mohn} as discussed below. In contrast, the F-induced state has a relatively larger band width of 0.61 eV [see Fig. 2(a)], thereby giving rise to the spin-unpolarized ground state.

The fact that the F-induced state is largely delocalized over C atoms of the ${\beta}$ sublattice leads us to speculate that the SIE inherent to the PBE functional may cause the incorrect prediction of the spin-unpolarized solution as a ground state. We note that the local or semi-local DFT scheme tends to overestimate the stability of delocalized states due to the SIE:~\cite{sie1,sie2,sie3} that is, the LDA or GGA functional within which an electron interacts with its own charge density unphysically lowers the energy of delocalized states, thereby incorrectly predict some materials to be metals rather than insulators. In order to correct the SIE, we use the hybrid HSE functional~\cite{hse} to perform the spin-unpolarized and spin-polarized calculations for an isolated F adatom on graphene. Our hybrid-DFT calculations predict a spin-polarized ground state with a magnetic moment of 0.97 ${\mu}_{\rm B}$ per adatom, which energetically favors over the spin-unpolarized configuration by 32 meV. Therefore, we can say that the HSE functional corrects the spurious delocalization to reproduce the observed~\cite{hong1,hong2,nair} local magnetic moment in fluorinated graphene.

The spin-polarized band structure of the F-graphene system obtained using the HSE calculation is displayed in Fig. 2(c). We find that (i) the F-induced state is almost fully spin-polarized to yield a magnetic moment of ${\sim}$1 ${\mu}_{\rm B}$ per adatom, (ii) the ${\pi}$ states of graphene have a band gap opening of 0.51 eV at the ${\Gamma}$ point, and (iii) the graphene states are spin-polarized with a spin splitting of less than 0.45 eV. Here, the spin polarization of the graphene states is attributed to their exchange interactions with the spin moment of the F-induced state. Consequently, the F-graphene system has a long-range spin polarization [Fig. 3(a)] not only due to the bipartite nature of the graphene lattice but also to the exchange spin-polarization effect in the valence states. Figure 3(c) shows that the calculated magnetic moment $m_i$ within the muffin-tin sphere centered at each C atom in the ${\alpha}$ (${\beta}$) sublattice gives a negative (positive) spin polarization. The values of $m_i$ are $-$0.01, $-$0.03, $-$0.02, and $-$0.02 (0.07, 0.03, and 0.02) ${\mu}_B$ for C$_0$, C$_2$, C$_4$, and C$_6$ (C$_1$, C$_3$, and C$_5$), respectively. Note that the magnetic moment within the muffin-tin sphere~\cite{mt} at the F adatom shows a positive spin polarization with $m$ = 0.06 ${\mu}_B$.

In Fig. 3(c), we also plot the values of $m_i$ in the H-graphene system obtained using the HSE and PBE calculations. We find that the HSE values of $m_i$ in the H-graphene system are similar to those in the F-graphene system, indicating that the magnitudes of exchange splitting in the H- and F-induced states are close to each other. It is notable that, for the H-graphene system, the PBE values of $m_i$ are relatively smaller than the corresponding HSE ones, caused by a relatively smaller exchange splitting of the H-induced state in the former compared to in the latter as discussed below. As a consequence, PBE gives rise to a relatively shorter range of spin polarization [see Fig. 3(b)] compared to the HSE result (Fig. 2 in the Supplemental Material).

\begin{figure}[tb]
\includegraphics[width=0.45\textwidth]{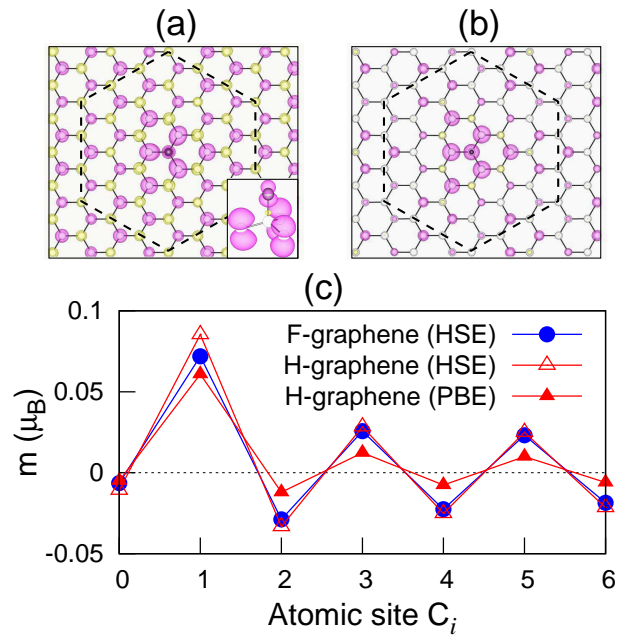}
\caption{(Color online) Spin densities of the (a) F-graphene and (b) H-graphene systems obtained using the HSE and PBE functionals, respectively. The majority (minority) spin density is displayed in dark (bright) color with an isosurface of 0.02($-$0.02) $e$/{\AA}$^3$. The inset in (a) enlarges the F-adsorbed region with the side view. For the F-graphene and H-graphene systems, the calculated magnetic moment $m_i$ within the muffin-tin sphere centered at the C$_i$ atom (see Fig. 1) is plotted in (c). Here, the muffin-tin sphere radius is taken as 0.71 {\AA}. }
\end{figure}

\begin{figure}[tu]
\includegraphics[width=0.45\textwidth]{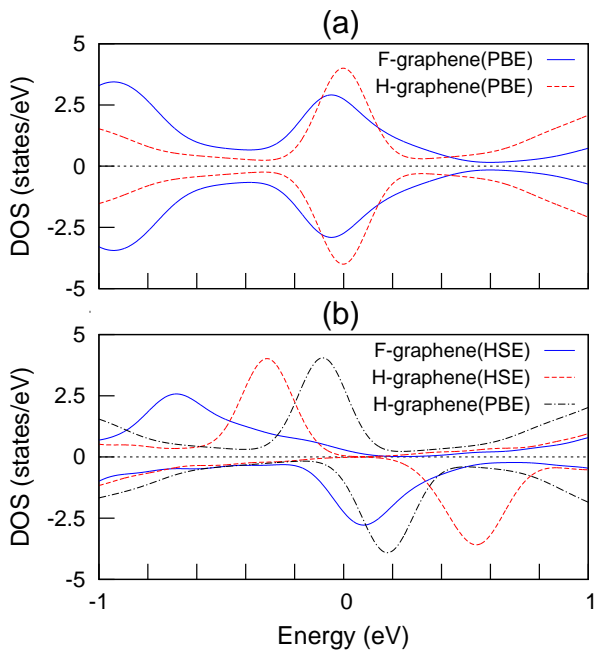}
\caption{(Color online) Calculated density of states for the (a) spin-unpolarized and (b) spin-polarized configurations of the F-graphene and H-graphene systems obtained using the PBE or HSE functional. The energy is relative to the Fermi level. The Gaussian broadening scheme with a width of 0.1 eV is used.}
\end{figure}

Figure 4(a) shows the density of states (DOS) for the spin-unpolarized configuration of the F-graphene system obtained using the PBE functional, together with the result of the H-graphene system. We find that the F-graphene system has the DOS of $D$ = 2.74 states/eV per spin at $E_{\rm F}$, which is smaller than the corresponding one ($D$ = 4.00 states/meV) for the H-graphene system. This reduction of $D$ in the F-graphene system does not allow for the presence of a ferromagnetic instability.~\cite{pbesto} On the other hand, the H-graphene system satisfies the Stoner criterion for the ferromagnetic instability: i.e., $D{\cdot}I$ $>$ 1, where the Stoner parameter $I$ can be estimated as ${\sim}$0.26 eV by the spin splitting of the H-induced state (see Fig. 4(b) and Fig. 1 in the Supplemental Material~\cite{supple}) obtained using PBE. Such a spin splitting of the H-induced state increases to ${\sim}$0.86 eV by the HSE calculation, as shown in Fig. 4(b). This increase of exchange splitting in HSE gives the relatively larger values of $m_i$ compared to the PBE values [see Fig. 3(c)]. The resulting HSE value of $I$ = 0.86 eV in the H-graphene system is slightly larger than the corresponding one ($I$ = 0.77 eV) in the F-graphene system, which is estimated by the spin splitting of the F-induced state [see Fig. 2(c) and Fig. 4(b)]. Thus, we can say that the HSE calculation for the F-graphene system fulfills the Stoner criterion with $D{\cdot}I$ = 1.99 $>$ 1 to yield a ferromagnetic ground state.~\cite{hsedos}

The present HSE result that an isolated F adatom induces a magnetic moment of 0.97 ${\mu}_B$ in graphene can be considered in reasonable agreement with the experimental observation of magnetic F adatoms with spin 1/2.~\cite{nair} It is, however, noticeable that the experiment~\cite{nair} did not detect a sign of ferromagnetic ordering but observed a purely paramagnetic behavior. Here, only one out of ${\sim}$1,000 F adatoms was estimated to contribute to the paramagnetism. This observed paramagnetism was explained by clustering of F adatoms, such that neighbouring adatoms residing on
different graphene sublattices do not contribute to the overall
magnetic moment. We note that the average spacing between paramagnetic centers located at cluster edges was estimated to be ${\sim}$10 nm,~\cite{nair} which is larger than the distance (1.3 nm) between neighboring F adatoms in our employed supercell. Thus, in order to measure a ferromagnetic ordering in the F-graphene samples, it would be required to reduce the spacing between magnetic moments either by decreasing the cluster sizes which may be controlled by the sizes of ripples in graphene~\cite{meyer} or by adsorbing F adatoms at a sufficiently low temperature to avoid the tendency towards clustering.


In summary, using a first-principles DFT calculation with the HSE exchange-correlation functional, we investigated the spin polarization for an isolated F adatom on graphene, which has not been adequately described by previous semilocal DFT calculations.~\cite{liu,santos,sofo2} The present hybrid DFT calculation showed that an isolated F adatom induces a spin moment of ${\sim}$1 ${\mu}_{\rm B}$ in graphene, contrasting with the semilocal DFT prediction of a nonmagnetic ground state. We found that F adsorption gives rise to a long-range spin polarization of $p_z$ orbitals in graphene due to the bipartite nature of the graphene lattice and the exchange spin-polarization effect in the graphene states. Thus, our prediction of the spin-polarized ground state in the F-graphene system provides theoretical support for the recent experimental observations~\cite{hong1,hong2,nair} of local magnetic moments in fluorinated graphene.

This work was supported by National Research Foundation of Korea (NRF) grant funded by the Korean Government (NRF-2011-0015754). The calculations were performed by KISTI supercomputing center through the strategic support program (KSC-2012-C3-36) for the supercomputing application research.

$^*$ Corresponding author; chojh@hanyang.ac.kr


\begin{thebibliography}{99}
\bibitem{mt} The muffin-tin sphere radius for the F atom was taken as 0.75 {\AA}.
\bibitem{pbesto} Assuming that the PBE value of $I$ in the F-graphene system is the same as that (0.26 eV) in the H-graphene system, PBE does not satisfy the Stoner criterion with $D{\cdot}I$ = 0.71 $<$ 1.
\end{thebibliography}
\end{document}